\documentstyle[bbm,amsfonts,epsfig,12pt,here]{article}

\newcommand {\eqref} [1] {(\ref {#1})}
\newcommand {\slsh} [1] {\not{\hbox{\kern-2pt${#1}$}}}

\newcommand{\drawsquare}[2]{\hbox{%
\rule{#2pt}{#1pt}\hskip-#2pt
\rule{#1pt}{#2pt}\hskip-#1pt
\rule[#1pt]{#1pt}{#2pt}}\rule[#1pt]{#2pt}{#2pt}\hskip-#2pt
\rule{#2pt}{#1pt}}

\newcommand{\Yfund}{\raisebox{-.5pt}{\drawsquare{6.5}{0.4}}}
\newcommand{\Yasymm}{\raisebox{-3.5pt}{\drawsquare{6.5}{0.4}}\hskip-6.9pt%
                     \raisebox{3pt}{\drawsquare{6.5}{0.4}}%
                    }
\newcommand{\Ysymm}{\Yfund\hskip-0.4pt%
                    \Yfund}
\def\symm{\Ysymm}

\def\drawbox#1#2{\hrule height#2pt
        \hbox{\vrule width#2pt height#1pt \kern#1pt
              \vrule width#2pt}
              \hrule height#2pt}

\def\Asym#1#2{\vcenter{\vbox{\drawbox{#1}{#2}
              \kern-#2pt       
              \drawbox{#1}{#2}}}}

\def\asymm{\Asym{6.4}{0.3}}

\def\basymm{\overline{\asymm}}

\def\lsim{\mathrel{\rlap{\lower3pt\hbox{\hskip0pt$\sim$}}
    \raise1pt\hbox{$<$}}}
\def\gsim{\mathrel{\rlap{\lower4pt\hbox{\hskip1pt$\sim$}}
    \raise1pt\hbox{$>$}}}

\newcommand {\beq} {\begin{equation}}
\newcommand {\eeq} {\end{equation}}
  \newcommand {\ber}{\begin{eqnarray*}}
  \newcommand {\eer} {\end{eqnarray*}}
\newcommand {\beqn}{\begin{eqnarray}}
  \newcommand {\eeqn} {\end{eqnarray}}

\newcommand{\Ntwo}{${\cal N}=2\ $}
\newcommand{\None}{${\cal N}=1\ $}

\def\Acknowledgements{\bigskip  \bigskip {\begin{center} \begin{large}
              \bf ACKNOWLEDGMENTS \end{large}\end{center}}}

\begin{document}

\begin{titlepage}

\begin{flushright}
{\em TO THE MEMORY OF IAN KOGAN}
\end{flushright}
\vskip 0.4cm
\begin{flushright}
{{\tiny CERN-TH/2003-123}

{\tiny
FTPI-MINN-03/15, UMN-TH-2203/03}}
\end{flushright}
\vskip 0.4cm
\centerline{\Large \bf  Remarks on Stable and Quasi-stable \boldmath{$k$}-Strings}
\centerline{\Large \bf  at Large \boldmath{$N$}}
\vskip 0.4cm
\centerline{\large A. Armoni ${}^a$ and M. Shifman ${}^{a,b}$}
\vskip 0.1cm
\centerline{${}^a$ Theory Division, CERN}
\centerline{CH-1211 Geneva 23, Switzerland}
\vskip 0.2cm
\centerline{${}^b$ William I. Fine Theoretical Physics Institute, 
University
of Minnesota,}
\centerline{Minneapolis, MN 55455, USA$^\star$}

\vskip 0.2cm

\begin{abstract}

We discuss $k$-strings in the large-$N$ Yang-Mills theory and its
supersymmetric extension.
 Whereas the tension of the {\em bona fide} (stable) QCD string is expected to
depend only on the $N$-ality of the representation, 
tensions  that depend on  specific representation $R$
are often
reported in the lattice literature.
 In particular, 
adjoint strings are discussed and  found in certain simulations.
We clarify this issue by 
systematically exploiting the notion of the
quasi-stable strings which becomes well-defined at large $N$.
The  quasi-stable strings with
representation-dependent tensions decay, but the decay rate
(per unit length per unit time)
is suppressed as
$ \Lambda^{2} {\cal F} (N)$ where ${\cal F} (N)$ falls off as a function of
$N$. It can be determined on the case-by-case basis.
The quasi-stable strings eventually decay into stable strings whose
tension indeed depends  only on the $N$-ality. 

We also briefly
review  large-$N$ arguments showing why the Casimir formula for the
string tension cannot be correct,  and present additional arguments in
favor of the sine formula. Finally, we comment on
the relevance of our estimates to Euclidean lattice measurements.

\end{abstract}

\vspace{0.3cm}

\noindent
\rule{2.4in}{0.25mm}\\
$^\star$ Permanent address.
\end{titlepage}

\section{Introduction}

\noindent

In confining theories, such as the Yang-Mills theory, non-supersymmetric 
or supersymmetric ({\None} gluodynamics),
heavy (probe) color sources
in the fundamental representation  are connected by color flux tubes,
fundamental QCD strings.
The fundamental string tension is of the order of
$\Lambda^2$ where $\Lambda$
is the dynamical scale parameter of the gauge theory under consideration,
and its transverse size is of the order of $\Lambda^{-1}$.
Both parameters are independent of the number of colors
(in what follows the gauge
group is assumed to be SU($N$)) and, besides $\Lambda$, can contain only
numerical factors.

Significant effort has been invested recently
in studies of the  flux tubes induced by color sources
in higher representations of SU($N$), mostly in connection
--- but not exclusively --- with high-precision lattice calculations,
see e.g. \cite{1,2,3,4,Forcrand,Deldar,Bali} 
and references therein. The composite flux tubes 
attached to such
color sources are also known as $k$-strings, where $k$ denotes the
$N$-ality of the color representation under consideration.
The $N$-ality of the representation with $\ell$ upper and
$m$ lower indices (i.e. $\ell$ fundamental
and $m$ anti-fundamental)  is defined as $k= |\ell -m|$.

Some qualitative features of the
$k$-strings are well understood. The most important feature is that
the string tension does not depend on the particular representation
of the probe color source, but only on its $N$-ality.
Indeed, the particular Young tableau of the representation
plays no role, since all representations with the given $N$-ality
can be converted into each other by
emitting an appropriate number of soft gluons. For instance,
the adjoint representation has zero $N$-ality; therefore,
the color source in the adjoint can be completely screened by gluons,
and the flux tube between the adjoint color sources should not
exist. The same is true for any representation with
$N$ fundamental or $N$ anti-fundamental indices. 
Theoretical ideas regarding confinement in lattice gauge theories
and the role of 
$N$-ality are extensively reviewed  in Ref.~\cite{Greensite:2003bk}.

At the same time, one can find in the literature  ---
especially, devoted to lattices --- reports on
the  adjoint string tension at intermediate distances, measurements
of distinct string tensions
for symmetric and anti-symmetric representations of one
and the same $N$-ality, and so on. For instance, in Ref.~\cite{Deldar}
which deals with SU(3) gauge group, the string tensions (at
distances up to  $\sim 2$ fm)
are
measured for representations \underline{8} 
of $N$-ality 0, \underline{3} and \underline{6}
of $N$-ality  2,  \underline{10}
of $N$-ality  3, and for some other representations.
The question is: ``Is there a conflict between the
above results and the general principles?"

A part of the apparent contradiction is due to
semantics. There is an objective difficulty too:
quantitative analysis is hindered by the fact that seemingly
there is no small
expansion parameter
in the $k$-string problem, that could play a guiding role.

In this paper we demonstrate that,
using $1/N$ as such a parameter, one can not only
resolve the above conflict, but put the analysis on a relatively
quantitative basis.
(Here $N$ is the number of colors). We assume the 't Hooft scaling,
i.e. $g^2 \sim 1/N$,  so that $\Lambda$ is kept fixed.

\begin{figure}[H]
  \begin{center}
\mbox{\kern-0.5cm
\epsfig{file=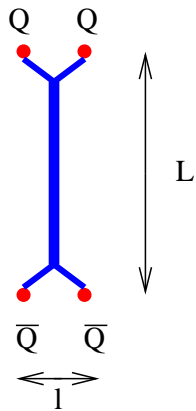,width=2.5true cm,angle=0}}
  \end{center}
\caption{A flux tube for the 2-string.}
\label{st}
\end{figure}

Gauge theories at strong coupling  have no
expansion parameter other than $1/N$. Within the 't Hooft
framework, the $1/N$ expansion is abstracted from the Feynman graphs.
However, since 
one in fact uses only the topology of the graph,
the result obtained is valid to all orders in the gauge coupling $Ng^2$
and is, thus, nonperturbative. The leading term in the
$1/N$ expansion corresponds to the planar 't Hooft graphs, the 
next-to-leading to the 't Hooft graphs with one handle, and so on.
This feature  --- the topological character of the
$1/N$ expansion --- is inherited and fully incorporated into string theory.
The string perturbation theory is the expansion in
$g_{\rm st}$. For closed strings it is $g_{\rm st}^2$
that enters, which corresponds to a $1/N^2$ expansion
in gauge theories. Non-perturbative stringy effects (D-branes),
that are exponential in $1/g_{\rm st}$, may give rise to 
exponential terms
of the type $\exp{(-N)}$ or $\exp{(-N^2)}$ in  gauge theories.

At large $N$
the problem of composite strings
becomes, in essence, a  weak
coupling problem. A crucial starting observation
is that at $N=\infty$ the fundamental strings (i.e. strings that
connect a quark-antiquark pair in the fundamental representation) 
become  free
and non-interacting, while their interaction emerges only
at the level of $1/N^2$ corrections \cite{5}.

In the remainder of this paper
we will adopt the following convenient convention:
the length of the strings under consideration
will be chosen
to be $N$-independent (but certainly
  much larger than $\Lambda^{-1}$).
The $N$ dependence is reserved
for  time duration of the measurement.

We view a $k$-string as a bound state of $k$ elementary fundamental
strings (see Fig. \ref{st}). This standpoint 
is natural in the gauge
theory/supergravity correspondence \cite{AdSCFT} where the $k$-string
world sheet (Wilson loop) is described as $k$-coincident elementary world sheets
\cite{Wilson-Mal,Gross:1998gk}. Accordingly, in   field theory one  can 
 {\em
define} the  $k$-string Wilson loop as $k$ coincident fundamental
Wilson loops. In doing so, one will 
deal with the probe source which
belongs to a {\em reducible} representation
with $k$ upper indices. If the (Euclidean)
observation time is sufficiently large, all excited-string contributions will
die out, and one will  measure the ground-state string tension.
Alternatively, for the study of quasi-stable strings,
one can single out a particular representation of the probe source, by considering
the Wilson loop in the appropriate representation.

The question we address below is the life-times of quasi-stable strings and
tension splittings. Our approach is based on a Minkowskian (quasiclassical)
picture of string interactions which is especially appropriate
for the description of various tunneling transitions. 
Some elements of this picture can be traced back to the 1970's
\cite{Nussinov}.
The validity of the quasiclassical approximation is justified by the smallness
of $1/N$. We then translate  our results in the 
Euclidean language appropriate for lattice simulations. 

Using general properties of weakly coupled
  systems we will show that the decay
rates of the quasi-stable strings (per unit length per unit time) scale as
$\Gamma \sim \Lambda^{2} {\cal F} (N)$. The dependence ${\cal F} (N)$ is either 
 $N^{-2}$ in the case of the adjoint string,
  $\exp (-N^2)$ is the case of a decay from a pure higher (i.e. non-anti-symmetric)
  representation into the ground state (the anti-symmetric
representation), or $\exp (-N)$ when $k \sim N$
(the saturation limit). The common reason behind the $N$ dependence is
that all these processes are non-planar in their nature and, therefore,
in the large-$N$ limit they will never occur (an infinite amount of
  time is needed to observe them). 

For  translating our results
in the Euclidean language it is important to understand
into which particular final state quasi-stable strings decay.
We will comment on this issue too. We expand our previous results
\cite{5} on the $k$-string tensions by further analyzing
string interactions at large/small distances. In particular, we explain, from a somewhat
new angle, the impossibility of the Casimir scaling\,\footnote{The very 
term {\em Casimir scaling}
was introduced in Ref.~\cite{1}.} for the {\em bona fide} 
$k$-strings\,\footnote{By  {\em bona fide} $k$-strings
we mean the stable strings with all $k$ indices antisymmetrized.} 
and what may mimic the Casimir scaling in lattice measurements
at intermediate distances.

The organization of the paper is as follows: Section 2 presents
short theoretical and lattice-data reviews on Wilson loops and
$k$-strings. In Sect. 3 we estimate the decay rates of quasi-stable
strings
in various situations. In Sect. 4 we introduce the notion of  
 the saturation 
limit and discuss the relevance of the sine formula in this limit.
Section 6 is devoted to oversatuarted strings. In Sect. 7 we provide
a physically transparent
 explanation of the interaction term $k^3/N^2$ in the expansion
of the sine formula. In Sect. 8 we introduce  and
discuss the  tension
deficit for composite strings. In 
Sect. 9 we suggest a
tentative explanation why the Casimir scaling could be observed at
intermediate distances. Section 10 is devoted to $k$-strings in the
``orientifold field theories''. In Sects. 11
we discuss the impact of our results on the Euclidean lattice calculations.
Finally,  Sect.  12 presents
 concluding comments and 
discussion of  relevance of our estimates to 
 $N=3$. A toy (exactly solvable)
model illustrating  $k^3/N^2$ in the expansion
of the sine formula in the tension can be found in the Appendix.

\section{Preliminaries: theoretical background, lattice data}
\label{prel}
  
The purpose of this section is to define and  review the notion
of stable (quasi-stable)
$k$-strings and review the way their tensions are measured. We also
briefly summarize  our previous paper \cite{5}.

\subsection{$k$-Strings --- theoretical background}
\label{kstb}

\noindent
Suppose  we wish to measure a long-distance quark-antiquark potential
in  pure SU$(N)$ gauge theory or in its minimal supersymmetric
extension, supersymmetric
gluodynamics. We assume that the quark-antiquark pair is in a specific
representation $R$. The expected long-distance potential is
\beq
V = \sigma L\,.
\eeq

The stable string tension $\sigma$ should not depend on the specific
representation $R$ of the quark-antiquark pair, but rather on its
$N$-ality $k$, since a soft gluon can transform a representation $R$
into a representation $R'$ within the same $N$-ality. For this reason
 we can evaluate the string
tension by considering a {\em reducible} representation with
a given $N$-ality $k$ instead of a specific representation $R$,
provided this reducible representation
contains $R$. For instance, one can consider, as a probe source,
an ensemble of $k$ heavy quarks (in the fundamental
representation)
nailed in close proximity from each other.
Another source, at distance $L$ from the former,
will be composed of  $k$ heavy anti-quarks. Since the fully 
anti-symmetric representation 
is expected to have the lowest
energy,  the string attached to a source
in a reducible representation
 will evolve into the anti-symmetric string 
after a certain time $\tau$.
The time $\tau$
is definitely  $\gsim$  than a typical inverse splitting
between the energies of the anti-symmetric string and other
representations. In fact, sometimes it may be  much larger.

By the same token, if the probe charges have $N$-ality zero,
even if they are connected by a string at time zero,
it will inevitably evolve into a no-string state since
such charges can be totally screened. For instance, the adjoint
probe heavy quark is screened by a gluon. Another example: if 
 one has $N$ fundamental quarks separated 
from $N$ fundamental anti-quarks by a large distance $L$
eventually each of the two $N$-quark ensembles
will develop string junctions (a baryon vertex), and there will be no
string connecting the two ensembles.

\begin{figure}[H]
  \begin{center}
\mbox{\kern-0.5cm
\epsfig{file=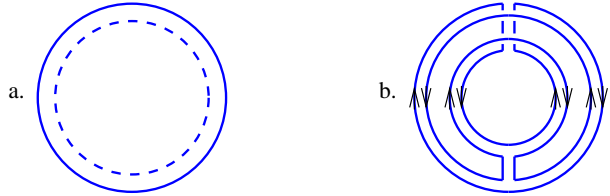,width=8.0true cm,angle=0}}
  \end{center}
\caption{(a). A screening of an adjoint Wilson loop by a dynamical
adjoint loop. (b). The same process described in 't Hooft double index
notation. The process in non-planar.}
\label{adjoint}
\end{figure}

A graphic illustration is presented in Fig.~\ref{adjoint}.
 Suppose that
one calculates the expectation value of a Wilson loop in the adjoint
representation (i.e. induced  by  an adjoint
probe ``quark''). Dynamically, a Wilson ``counter-loop'' in the adjoint
representation will be formed thus screening the
non-dynamical probe quark. In physical terms,
a gluon lump is produced which combines with the probe adjoint source
to form a color singlet. 

The fundamental string, by definition,
 is the one that connects a fundamental
heavy quark with an anti-quark.
The interaction between the fundamental
strings is via the glueball exchanges. The process
is {\em non-perturbative} in the  coupling $g^2 N$.
Although one uses the Feynman graphs in the 't Hooft
representation in order to analyze the $N$ dependence,
by no means it is implied that the results thus obtained
are perturbative in the gauge coupling and ``correspond to few-gluon
exchange.'' As is standard in the 't Hooft framework,
since the $1/N$ analysis is based only on topology
of the graphs (e.g. planarity vs. non-planarity),
the results are expected to fully represent the non-perturbative
strong coupling gauge dynamics. 
 
\begin{figure}[H]
  \begin{center}
\mbox{\kern-0.5cm
\epsfig{file=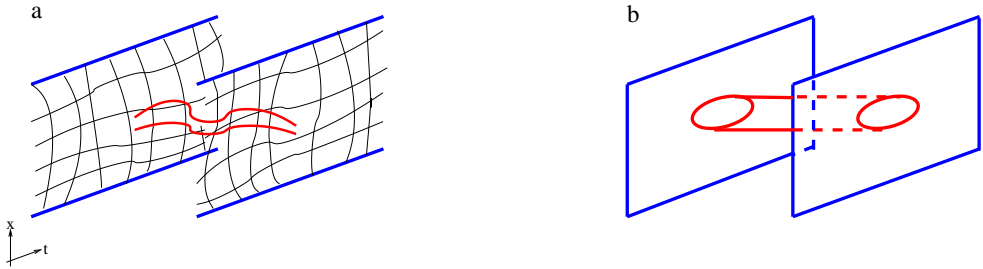,width=13.0true cm,angle=0}}
  \end{center}
\caption{The interaction of two ``fundamental'' strings: (a)
Field-theory picture --- (two)gluon exchange; (b) String-theory picture ---
exchange of a closed string between two world sheets.}
\label{interaction}
\end{figure}

Since we deal with the interactions between color-singlet objects 
 and the interaction is non-planar, it
will be controlled by  $O(1/N^2)$, see Ref.~\cite{5}. At $N=\infty$ 
the interaction vanishes, and we
have  $k$ free fundamental strings with $\sigma_k = k\sigma_{\rm f}$. 
A sample  $1/N^2$ interaction between two fundamental Wilson loop is depicted
in Fig. \ref{interaction}a.
The fact that the corrections to the free strings relation
  $\sigma _k = k \sigma_1$ run in powers of $1/N^2$ was
used \cite{5} to exclude the Casimir scaling hypothesis.
On general grounds one can assert that at distances $\gsim \Lambda^{-1}$
the fundamental strings attract each other \cite{5}, while at
short distances there is a repulsion (see the end
of this section),  so that 
composite $k$-strings develop.

This picture is supported by string theory. String theorists 
prefer to call the Wilson loop a string world sheet. The realization
of the $k$-string will be simply $k$
coincident elementary (fundamental) strings,
or more precisely, a bound
state of $k$ elementary string world sheets,
see Fig. \ref{interaction}b. 

Within the AdS/CFT correspondence \cite{AdSCFT} we can elevate this model to a
quantitative level \cite{Wilson-Mal,Gross:1998gk}. Indeed, the
AdS/CFT correspondence is a
 natural framework for calculating the
$k$-string tension, since in the dual string picture we work
 at strong 't Hooft coupling. In addition, $1/N$ is represented by $g_{\rm st}$. 
In this framework the Wilson loop
on the AdS boundary is described by a fundamental world sheet that
extends inside the bulk AdS \cite{Wilson-Mal}. The value of the Wilson
loop is the area of the minimal surface, given simply 
by the Nambu-Goto action
\beq
\langle{\cal W}\rangle=\exp\left( -S_{\rm NG}\right).
\eeq
 Similarly, the $k$-string
Wilson loop \eqref{kstring} is described by  $k$ coincident world sheets.
In the supergravity approximation, $g_{\rm st}=0$, we obtain $\sigma _k =
k \sigma_1$ \cite{Gross:1998gk}, as expected. In order to calculate
$1/N$ corrections, one has to consider string interactions, namely to
go beyond the lowest order in $g_{\rm st}$. It is clear, however, that in the closed string
theory (and, in particular, in type IIB) the expansion parameter is
$g_{\rm st}^2=1/N^2$. 

The fact that the interaction between the fundamental strings
is proportional to $1/N^2$ and, thus, vanishes at
$N\to\infty$ is explained in detail in Ref.\cite{5}
Moreover, from the same work we know that
at distances $\gsim\Lambda^{-1}$ the interaction is attractive.
An  attractive potential between the fundamental strings
is also obtained \cite{DelDebbio:1992yp} in lattice
 strong-coupling calculations. What is the nature of the inter-string interaction at distances of the
order of the fundamental string thickness?

Logically there are two options. Either the parallel strings attract
at all distances (then $k$ closely situated 1-strings will glue together
forming a structureless flux tube  carrying $k$ units of $N$-ality),
or the attraction gives place to repulsion  at shorter distances
(then the $k$-string will have a substructure in the transverse
plane reminiscent of that of a nucleus). The large-$N$ expansion
of the $k$-string tension suggests that it is the latter option that
is realized in QCD. Indeed, on general grounds, with no
model dependence, one can show that at $N\gg 1$ and $k \gg 1$
$$
\left(\frac{\sigma_k}{k\sigma_{\rm f}} -1 \right) \propto \frac{k^2}{N^2}
$$
(see Sect. 4.1). The above $k$ dependence of the binding ``energy"
has no natural explanation in the picture of a forced ``collectivization"
inevitable under the assumption that the
parallel 1-strings attract at all distances. At least, we are unaware of
such an explanation
and were unable to  obtain it, in spite of several attempts.

At the same time,
the repulsion at shorts distances, that naturally leads to a nucleus-type
structure of the slice of the $k$-string, {\em and } explains the above
$k$ dependence of the binding ``energy" (see Sect. 7)
is an immediate consequence of the following consideration.

The trace of the energy-momentum tensor $\theta_\mu^\mu$ in pure
Yang-Mills
theory or in QCD with
massless fermions has the form (the scale anomaly)
$$
\theta_\mu^\mu =\frac{-b}{32\pi^2} G_{\mu\nu}^a\, G^{\mu\nu\,,\,\, a}
$$
where $G_{\mu\nu}^a$ is the {\em operator}
of the gluon field strength tensor ($b$ is the
first coefficient of the Gell-Mann--Low function). One can use $\theta_\mu^\mu (x)$
as a local probe of the energy density. Indeed the value of 
$\theta_\mu^\mu \propto G^2$, in the presence of a Wilson loop $W(C)$, 
measures the string tension
\cite{Novikov:xj},
\beq
\langle W(C) \rangle ^{-1} \,
 \int d^3 x\, \langle \theta_\mu^\mu (x)\,,\, W (C)\rangle_{\rm c,\,\,eucl}  = 2 \sigma R\,,
\eeq
where the subscript c stands for connected part.
The chromomagnetic part is
presumably negligible for the static flux tubes
attached to static color sources. This implies that
the energy density is proportional to the
expectation value of the operator $\vec E^2 (x)$, where
$\vec E$ is the chromoelectric field, and the arrow is used to represent
both vectorial and color indices. Then the string tension
(the energy per unit length in the $z$ direction)
is given by the integral
$$
\sigma \propto \int d^2 x\, \langle \vec E^2 (x)\rangle\,,
$$
where the integral runs in the perpendicular plane.

\begin{figure}[H]
  \begin{center}
\mbox{\kern-0.5cm
\epsfig{file=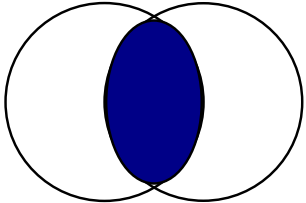,width=4.0true cm,angle=0}}
  \end{center}
\caption{The spatial slice of
two overlapping fundamental strings.}
\label{ofs}
\end{figure}

Two overlapping fundamental flux tubes are depicted in Fig. \ref{ofs}.
Now
$$
\sigma_2 \propto \int d^2 x\, \left\langle\left( \vec E_{{\rm f}\,1} (x)
+ \vec E_{{\rm f}\, 2} (x)\right)^2\right \rangle
= 2\, \sigma_{\rm f} + 2 \,\int d^2 x\, \left\langle  \vec E_{{\rm f}\,1}
(x)
\, \vec E_{{\rm f}\, 2} (x) \right \rangle
$$
The fluxes are fixed by the given (static) color sources.
This implies that the interference term is positive. It is not difficult
to see that it is also  suppressed by $1/N^2$, because generically
$\vec E_{{\rm f}\,1}$ is orthogonal to $\vec E_{{\rm f}\,2}$
in the color space; only the Cartan components
of the chromoelectric field are important. Thus, if two
parallel 1-strings overlap, the energy per unit length is larger than
$2\, \sigma_{\rm f} $ by $\Lambda^2/N^2$,
i.e. overlapping flux tubes repel each other.

Certainly, the arguments
presented above are somewhat quasiclassical.
We believe, however, that they are qualitatively
correct; they certainly lead to a self-consistent overall picture.
For further discussion see Sect. 7.

\subsection{A mini-review of the lattice literature on $k$-strings}
\label{mrllks}

\noindent
 The Euclidean/lattice  formulation of the problem is usually
given through the expectation value of a rectangular Wilson loop, 
\beq
\left\langle {\cal W}(C)
\right\rangle = \left\langle {\rm tr} \exp (i\int A _\mu ^a T^a_R\,  dx^\mu)
\right\rangle ,
\label{willo}
\eeq
where $T^a_R$ are the generators of SU$(N)$ in the given representation,
which may or may not be irreducible.
Sometimes, for practical purposes, it is more convenient
to consider a correlation function of two Polyakov lines. Assume that the 
(Euclidean) time direction is compactified, and one defines
two Polyakov lines in the time direction, separated by a distance $L$.
The Polyakov line is defined through the integral similar to 
(\ref{willo}), with the generators in the given representation $R$.
If time interval is large enough, the measurement of the
correlation function of two Polyakov lines must yield $\exp (-\sigma_R
L\, T)$.

The generators of the reducible representation with $N$-ality $k$ are given
by  tensor products of the fundamental representation.
For example, in the case of $N$-ality $2$ we have
\beq
\Yfund \otimes \Yfund = \Yasymm + \Ysymm \,.
\eeq
Therefore we can evaluate the $k$-string tension by considering the
following Wilson loop 
\beq
\left\langle {\cal W}_{\mbox{$ k$-string}}(C)
\right\rangle \equiv \left\langle\left[  {\cal W}_{\mbox{ fund}} (C)\right]^k
\right\rangle \,. \label{kstring}
\eeq
Physically, the above definition \eqref{kstring} represents
$k$-fundamental coincident Wilson loops. 
It is instructive to think of it as of
$k$ fundamental probe sources placed at one point.
The resulting composite probe source
is in the reducible representation of the
$N$-ality $k$, which includes a
mixture of irreducible representations, starting from fully anti-symmetric,
up to fully symmetric.

One of the main tasks of this paper is to bridge a gap between
theoretical studies of $k$-strings from the string theory side and
large-$N$ field theory side on the one hand, and lattice
studies, on the other hand. 
Therefore, it is natural to give a brief
  summary of 
 some lattice  works devoted to   $k$-strings. 

Let us start with analytic studies, namely the strong coupling
expansion.  Since the seminal work of Wilson \cite{Wilson:1974sk} it is known
that in the Yang-Mills  theory with no dynamical matter in the fundamental
representation, the expectation value of a {\em large} Wilson loop induced by
a heavy (non-dynamical) 
fundamental quark will trivially exhibit an  area law
\beq
\left\langle {\rm tr}\, \exp\,\left(  i\int \, A_\mu dx^\mu
\right)\right\rangle  \sim \exp \left( - \sigma\, {\cal A}\right).
\label{ququ}
\eeq
Here $\sigma$ the string tension, $\sigma \sim (\ln g^2) a^{-2}$,
where $a$ is the lattice cite and $g$ is the gauge coupling, $g\to\infty$.

The area law for the fundamental Wilson loop
is also  well established  in the {\em continuum limit}, through
numerical simulations. To detect the area law
there is an obvious necessary condition on the
area of the loop, 
\beq
{\cal A} \gg \Lambda^{-2}\,.
\label{cond}
\eeq
The slope in front of the area, Eq. (\ref{ququ}) ---
the fundamental string tension --- is measured to a reasonable accuracy.
Moreover, a perpendicular slice of the fundamental
string was studied too. A typical transverse dimension
of the fundamental string is $\sim 0.7$ fm.

On the other hand, an area law has been also detected for the adjoint Wilson
 loop for contours 
satisfying the condition (\ref{cond}).  For instance, typical
distances in Ref.~\cite{Deldar} were $\sim 1.5$ to 2 fm.
The adjoint string tension was measured. The  reported ratio
$\sigma_{\rm adj}/\sigma_{\rm f}$ is close
to the Casimir formula which for SU(3) yields 9/4.
(Please, note that this number is larger than 2; we will return to this point
in Sect.~\ref{mocs}.)
In fact, the adjoint strings emerge practically
in all lattice studies.
So far, only one work \cite{Forcrand} reports an observation
of the adjoint string breaking.\footnote{This work is based
on a technique different from all other calculations.}

A linear
potential between an adjoint probe quark and anti-quark 
implies the existence of 
 a {\em quasi-stable string}. It is certain that for
asymptotically large  areas, the area law {\em  must}
 give place to the perimeter one;
the linear potential {\em  must} flatten off at the level
 corresponding to the
creation of two gluelumps \cite{Forcrand}. The effect is not seen, however,
in the existing simulations. The
 question  to ask  is: ``what is the critical size/time
needed for detecting the adjoint string breaking?''

Now, the very same dynamical adjoint ``counter-loop'' as in Fig.~\ref{adjoint}
(or a few ``counter-loops'')
can transform external probe quarks 
in a given representation $R$ to a different representation $R'$
with the same $N$-ality. 
In physical terms, if we have, say,  an anti-symmetric source $Q^{[ij]}$,
we can convert it into an object with two symmetric indices $\{ij\}$ by adding
a soft gluon.
It is clear, then, why one expects  the $k$-string tension to depend
on the $N$-ality, and not on the specific representation $R$.
The genuine stable string for the given $N$-ality is expected to be attached
 to the probe source in the anti-symmetric representation.
The probe sources in other representations
with $k$ (upper) indices give rise to quasi-stable strings
whose tension is supposed to depend not only on the $N$-ality,
but on the particular representation. If the probe sources
have $k>[N]/2$, we will call such strings {\em oversaturated}.
Representations which have both upper and lower indices
can be viewed as bound states of $k$-strings and some number of adjoint strings.
Obviously, they are also quasi-stable. We will refer to such strings as
{\em bicomposites}.

Lattice measurements
of the Wilson loops or Polyakov line correlators seem to defy the
above argument. They
yield string tensions which depend on the particular representation 
under consideration rather than on 
the $N$-ality of the representation. For instance, in Ref.~\cite{3}
which treats the SU(4) and SU(6) cases
and measures the anti-symmetric and symmetric Wilson loops, distinct tensions are
obtained for the anti-symmetric and symmetric
two-index representations.
 It was
argued \cite{3} that the symmetric string  is not stable and that it
will have to decay into two fundamental strings. In light of the above,
again, the most crucial question
is what contour sizes are needed to
exhibit the decay of the symmetric string into anti-symmetric.
It is clear that the answer depends on the decay rate 
of the symmetric string.

Returning to the tensions of the stable (anti-symmetric)
strings, there
is no consensus in the lattice literature on the $k$-string
tension. The  Casimir scaling  is often reported.  
In \cite{2} is it argued that in three dimensions the $k$-string tension is 
very close to the  Casimir scaling. A similar claim is made in
 \cite{Bali} for four-dimensional SU(3) theories, where the Casimir behavior
is found for symmetric representations.
On the other hand, there are dedicated studies \cite{3}
which favor the sine formula for
the antisymmetric strings.

As we have already mentioned,
the  Casimir scaling behavior for a stable string
\cite{2} is
in contradiction with our analytical considerations \cite{5}:
the string tension should be expandable in powers of $1/N^2$ rather than
 $1/N$.

\section{Decay rates of the quasi-stable strings: case study}
\label{dpqss}

In this section we will give systematic
estimates of the quasi-stable string decay rates.
The strings  are assumed to live in the Minkowski time.
As was mentioned, the length of the strings under consideration
will be chosen
to be $N$-independent (but certainly
  much larger than $\Lambda^{-1}$).
The $N$ dependence is reserved
for  time duration. We will calculate the $N$ and representation
dependence of the
decay rates per unit time per unit length 
of the string. 
We will assume that $N\gg 1$ to infer
 regularities in this limit,
and then speculate as to which extent 
 our results may survive  extrapolation to $N=3$.
 The question as to how these rates
are reflected in the lattice measurements
of the Wilson loops/Polyakov lines 
is deferred till Sect.~\ref{impli}.

\subsection{Fundamental string in QCD with dynamical quarks}
\label{fsiqcdwdq}

\noindent

To warm up, let us start our consideration
from SU$(N)$ Yang-Mills theory with, say, one dynamical
(Dirac) quark in the fundamental representation. In this context,
the title of this section might seem provocative, since, as everybody 
knows, dynamical quarks screen the fundamental probe color charges,
and the string does not exist in this case.

Remember, however, that at large $N$ the dynamical quark production is
suppressed by $1/N$. The quark loops decouple at $N=\infty$;
in this limit the fundamental stable string exists, and its tension
is well-defined and equal to a numerical constant times $\Lambda^2$.
At large but finite $N$ the string is quasi-stable. The probability of 
its breaking
per unit length per unit time was found long ago~\cite{Nussinov},
\beq
\Gamma_{\rm f} \sim   \Lambda^2\, N^{-1}\,.
\label{ltos}
\eeq
This follows from the fact that planar graphs with one quark-loop
insertion are suppressed by the factor $N^{-1}$.

After the string is broken by the dynamical quark-antiquark
pair it decays into two strings --- one attached to the pair
$Q\bar q$, another to $\bar Q q$ (here $Q$ is the probe non-dynamical quark, 
while $q$ is a dynamical one). Let us have a closer look
at the final state (the reason why we dwell on this issue will 
become clear later). 

If we limit ourselves to the classical approximation,
we will have to conclude that the final-state strings are typically long,
so that the overall energy carried by these two strings is
$\sigma_{\rm f}L$ mod $\Lambda$,
i.e. the same as the energy  stored in the original string before
breaking. Indeed, the $\bar q q$ pair is produced from the vacuum
in a ``soft" manner, locally, at a certain point on the string.
Typically, the momenta of $\bar q$ and $q$
at the time of production are of the order of $\Lambda$.
Larger momenta are highly unprobable since
they can be obtained only from exponentially suppressed
tails of the ``wave function."

In the large $N$ limit, the configuration obtained after the
production of the $\bar q q$ pair evolves in time with no further
breakings. Let us have a closer look at the ``halves" of the broken
string.
Each consists of two well separated quarks (say, $\bar Q q$)
connected by a long well developed string carrying energy scaling as $L$.
The probability for this configuration to materialize as a ground-state
meson is exponentially suppressed. In typical resonances, strings have
sizes $\sim \Lambda ^{-1}$ (stricktly speaking, they cannot be called
strings, rather ``sausages"). A long string resides in
  exponentially suppressed tails of the meson wave functions.

On the other hand, exactly this configuration ---
two   quarks connected by a long well developed string ---
is typical for highly excited meson states (radial excitations).
Being highly excited, they can be
treated quasiclassically. In such mesons the light quark
is ``smeared" around the heavy one at a large distance,
which parametrically grows  with the excitation number.
The energy stored in the gluonic degrees of freedom
is proportional to this distance.
If it were not for the $1/N$ suppression,
this meson would decay into the ground-state
$\bar Q q$ meson plus multiple glueballs.

\begin{figure}[H]
  \begin{center}
\mbox{\kern-0.5cm
\epsfig{file=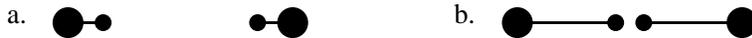,width=10.0true cm,angle=0}}
  \end{center}
\caption{a. A dynamical pair of soft quarks 
near the external charges. b. Creation
of a dynamical pair from the vacuum in the process of the string breaking.
The dynamical quarks are hard.}
\label{avp}
\end{figure}

As was mentioned, quantum-mechanically, there is a certain overlap of the final state described above
with the ground state mesons  $Q\bar q$ and $\bar Q q$ depicted in Fig.~\ref{avp}a. This
projection must be small, however.
The smallness is unrelated to $N$; it is controlled
by the parameter $L\Lambda$. We will return to the issue of the
smallness of this projection in Sect.~\ref{impli}.

Equation (\ref{ltos}) implies a particular relation between the
mass of a highly excited meson and its decay width, namely,
\beq
\Gamma_n = C \, M_n\, N^{-1}\,,\quad C\sim 0.5\,.
\label{maw}
\eeq
Here $n$ is the excitation number, $n\gg 1$. The above relation was numerically studied in the
't Hooft model and was confirmed to be valid \cite{bs}.

\subsection{Adjoint string}
\label{as}

\noindent
Now, consider pure SU$(N)$ Yang-Mills theory (no dynamical
quarks) with the probe heavy sources in the adjoint representation.
The screening of the sources occurs via  creation of a pair of gluons.
At the hadronic level, the string breaking
is equivalent to the statement that the operator $\bar QQ$
produces a pair of (color-singlet) mesons of the type
$QG$. Here $Q$ is the field of the probe heavy quark
while $G$ stands for the gluon.
It is easy to see that the amplitude of this
process is suppressed by $1/N$, while the probability
of the string breaking (per unit length per unit time) is
$\Lambda^2 /N^2$ (see Fig. \ref{adjoint}b),
\beq
\Gamma_{\rm adj}   \sim  \Lambda^2\, N^{-2}\,.
\label{aslt}
\eeq
  The adjoint string is quasi-stable,
much in the same way as the one in Sect. \ref{fsiqcdwdq}.
The difference is that now the decay rate
is suppressed by $N^{-2}$ rather than $N^{-1}$. Although the distinction is
quantitative rather than qualitative,
it may entail consequences for numerical simulations.
Indeed, for  large $N$,
the available lattice sizes may be insufficient to
place Wilson contours large enough
to observe the adjoint string decay (see, however, Ref.~\cite{Forcrand}).

What can be said about the adjoint string tension
$\sigma_{\rm adj}$? 
On general grounds
\beq
\sigma_{\rm adj} =2 \,\sigma_{\rm f}\left( 1+ O\left( N^{-2}\right)\right)\,.
\label{aste}
\eeq
Since at distances $\gsim \Lambda^{-1}$ there is an attraction between
the fundamental strings, the $O\left( N^{-2}\right)$ term
on the right-hand side must be negative. To measure this term in the 
Euclidean lattice simulations one
would need to have the contour area ${\cal A} > \Lambda^{-2} \, N^2$.
However, the adjoint string will decay at parametrically smaller
areas, see Sect.~\ref{dases}. Therefore,
at asymptotically  large $N$ the adjoint string tension
is not measurable through existing lattice methods.
At $N=3$ the splitting between $\sigma_{\rm adj}$
and $2 \sigma_{\rm f}$ may be measurable, though,
if the interplay of numerical factors is favorable.

\subsection{Bicomposites}
\label{bicomp}

\noindent

In this section we will discuss quasi-stable strings combining features
of the adjoint strings and $k$-strings, namely, we will consider
the probe source of the type
\beq
Q^{i_1 ... i_k i_{k+1}}_j\,.
\label{combi}
\eeq
One can either deal with the reducible representation or pick up an irreducible one,
say, by anti-symmetrizing all upper indices.
In fact, the situation is very similar to that discussed in Sect.
\ref{as}, as long as $k$ does not scale with $N$.

In this case the (quasi-stable) string tension does {\em not} depend
on the $N$-ality $k$, but, rather, on the index
$\kappa$,
\beq
\kappa = n + m\,,
\label{kap}
\eeq
where $n$ is the number of fundamental
and $m$ anti-fundamental indices of the probe field $Q$.
For the probe source (\ref{combi})  $\kappa =k+2$.

The tension of the given bicomposite is 
\beq
\sigma_{\rm bicomp} =(2+k) \,\sigma_{\rm f}\left( 1+ O\left( N^{-2}\right)\right)\,.
\eeq
It decays into a $k$-string through the gluelump creation, Fig. \ref{adjoint}.
The decay rate is given by Eq. (\ref{aslt}) and is suppressed
by $1/N^2$. Since the binding energy is $O(1/N^2)$ and the 
critical contour area  scales
as $\ln (N^2/C)$, the determination of the binding energy of 
this quasi-stable string is impossible at asymptotically large $N$. 

As $k$ grows, on combinatorial grounds
one expects the decay rate  to increase,
and loose its $1/N^2$ suppression when $k$ becomes of the order of
$N/2$.

\subsection{\boldmath{$k$}-Strings and their excitations}
\label{ks}

\noindent

In this section we will consider
probe color sources of the type
$Q^{i_1,i_2,..., i_k}$ or $Q_{i_1,i_2,..., i_k}$,
with $k$ fundamental or anti-fundamental
indices. 
The values of $k$ to be discussed in this section
are of the order of $N^0$. The values of $k\sim N^1$
are considered in Sects. \ref{ksslsf} and \ref{dqsssl}.

In the general case, this is a reducible
representation of SU($N$).
To single out irreducible representations,
one must perform symmetrization (anti-symmetrization)
according to the given Young tableau.
Not to overburden the subsequent discussion,
for the time being we will limit ourselves to $k=2$
(higher values of $N$-ality will be considered in Sects.~\ref{kssl} and \ref{dqsssl}).
In this case there are two Young tableaux ---
full symmetrization and full anti-symmetrization, $Q^{\{ij\}}$ and 
$Q^{[ij]}$.
One can view the color source $Q^{ij}$ as the
source composed of two fundamental heavy quarks,
$Q^i$ and $Q^j$.
If $Q^i$ and $Q^j$
are placed close to each other, a flux tube which connects them with
two corresponding anti-quarks is called $2$-string, see Fig. \ref{st}.
In the general case of $k$ fundamental quarks, we would be dealing with
the $k$-strings.

In Ref.~\cite{5} we presented arguments that at distances $\gsim \Lambda^{-1}$
in the transverse direction the fundamental strings attract each other,
that the lowest-energy state belongs to the anti-symmetric string,
and that the tension splitting between the symmetric quasi-stable
string and anti-symmetric string scales as $\Lambda^2\, N^{-2}$. Numerical 
evidence also suggests 
(see \cite{1} and \cite{3})
that the symmetric quasi-stable 2-string has a larger tension than the 
anti-symmetric 2-string. 

If one studies the probe charges in the reducible representation, as in Fig. \ref{st},
containing a mixture of the symmetric and anti-symmetric
quasi-stable 2-strings, it is quite obvious that one will need
the time interval $\tau \gg N^2\,\Lambda^{-1}$ to resolve the tension splitting.
In the Euclidean language, at $T \gg  N^2\,\Lambda^{-1}$,
the excited (symmetric) string contributions dies off.

\vspace{1mm}

A question which is more subtle and interesting is as follows:
``Assume, one starts from the pure (excited) state, i.e.
one considers the Wilson loop for  $Q^{\{ij\}}$;
then what is its decay time?''

We want to show that the symmetric string does not decay
into the anti-symmetric one to any finite order in 
$1/N^2$. The decay rate is exponential.
Indeed, in order to convert the symmetric color representation into
anti-symmetric one has to produce a pair of gluons.
This takes energy of the order of $\Lambda$.
However, the string is not entirely broken,
rather it is restructured, with the tension splitting
$\sim \Lambda^2 N^{-2}$. To collect enough energy,
the gluon creation should take place not locally, but, rather
at the interval of the length $\sim \Lambda^{-1}\, N^2$.
This is then a typical tunneling process.

\begin{figure}[H]
  \begin{center}
\mbox{\kern-0.5cm
\epsfig{file=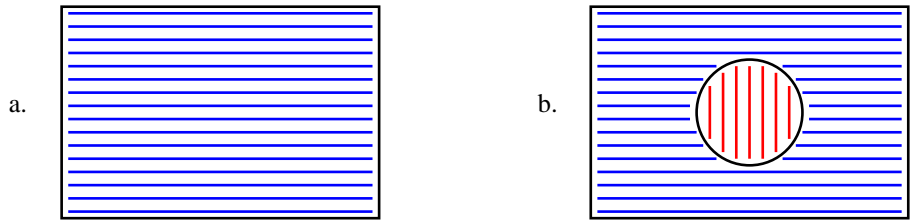,width=12.0true cm,angle=0}}
  \end{center}
\caption{The world-sheet of a 2-string. (a). A purely symmetric string
(denoted by horizontal lines). (b). The decay into an antisymmetric
string (vertical lines) via an expanding bubble.}
\label{symm}
\end{figure}

In fact, the decay rate can be found quasiclassically.
The world sheet of the symmetric string is shown in Fig. \ref{symm}a.
Its decay proceeds via a bubble creation, see Fig. \ref{symm}b.
The world sheet of the symmetric string is two-dimensional 
``false vacuum,'' while inside the bubble
we have a ``true vacuum,'' i.e. the surface
spanned by the anti-symmetric string. The tension difference
--- in the false vacuum decay problem, the vacuum energy difference ---
is ${\cal E} \sim  \Lambda^2 N^{-2}$, while the energy $T$
of the bubble boundary
(per unit length) is $T\sim \Lambda$. This means that the thin wall approximation
\cite{Kobzarev:1974cp,Voloshin:1995in}
is applicable, and the decay rate
$\Gamma$ (per unit length of the string per unit time)
is \cite{Kobzarev:1974cp,Voloshin:1995in}
\beq
\Gamma_{\symm\,\, \to\, \asymm} \sim \Lambda^2 \exp\left(-\frac{\pi\, T^2}{{\cal E}}\right)  
\sim \Lambda^2 \exp\left(-\gamma \, N^2 \right)\,,
\label{expdec} 
\eeq
where $\gamma$ is a positive constant of the order of unity.
Once the true vacuum domain is created through tunneling,
it will expand in the real time pushing the boundaries
(i.e. the positions of the gluons responsible for the conversion)
toward the string ends.

\section{\boldmath{$k$}-Strings: saturation limit and the sine formula}
\label{ksslsf}

The strings attached to the sources which have
$k$ fundamental indices (with no anti-fundamental), or {\em vice versa}
are generically referred to as $k$-strings. The stable $k$-strings (or {\em bona fide}
$k$-strings) are those attached to the fully anti-symmetric sources.
Others are quasi-stable $k$-strings. In this section we will assume that
$k$ scales as $N^1$ at large $N$.

\subsection{Saturation limit: generalities}
\label{kssl}

\noindent
In this section 
we will address the issue of the
stable $k$-string tension in
a limit which we call {\em saturation},
\beq
N\to \infty\,, \quad k\to \infty\,,\quad \frac{\pi k}{N}\equiv x\quad\mbox{fixed}\,.
\label{satli}
\eeq
We will combine the knowledge obtained from (i) symmetry properties;
(ii) the saturation limit; (iii) $N\to\infty$ limit with $k$ fixed;
and (iv) $1/N^2$ expansion to prove that the general formula for the 
$k$-string tension in the saturation limit is as follows:
\beq
\frac{\sigma_k}{k} = \sigma_{\rm f}\, f(x)\,,
\label{genfo}
\eeq
where the function $f$ can be expanded in {\em even} powers of $x$,
\beq
f(x) = 1 +a_1 \, x^2 + a_2 \, x^4+ ...
\label{epox}
\eeq
We then comment on how this peculiar $k$ dependence can be physically understood.

The central element of the proof is the $Z_N$ symmetry,
i.e. the fact that the {\em bona fide}
string tension $\sigma_k$ does not change under the replacements
\beq
k\to N-k\,,\quad\mbox{and}\quad k\to k+N\,.
\label{krepl}
\eeq
The function $f$ is dimensionless and, generally speaking, depends 
on two parameters, $k$ and $N$. Equation (\ref{krepl})
implies that $k$ may enter only in a very special way, namely,
through powers of $|\sin x |$. Thus, in general one can write
\beq
\sigma_k  =\frac{\sigma_{\rm f}}{\pi} 
\left[ c_1 (N) \, |\sin x | + c_2 (N) \, |\sin x |^2+ c_3 (N) \, |\sin x |^3 + ...\right]
,
\label{fexp}
\eeq
where $c_\ell (N)$ are coefficients.

Now, we make use of the facts that 
(a) the saturation limit should be smooth,
and (b) for fixed $k$ the expansion parameter
is $N^{-2}$ rather than $N^{-1}$.
Then  we conclude that  the coefficients
$c_{2\ell +1}= O(N^{1})$ while $c_{2\ell +2}= O(N^{0})$  (here $\ell = 0,1,2,...$).
Moreover, from the condition (iii) above it follows that $c_1 (N) = N$.

Omitting terms vanishing in the saturation limit,
we get 
\beq
\sigma_k  =\frac{\sigma_{\rm f}}{\pi}\, N\,  \left\{\,   |\sin x | + \,\, 
C_3\,  |\sin x |^3 + ...\right\} ,
\label{fexpp}
\eeq
which is equivalent to Eq. (\ref{epox}) (here $C_3$ 
and possibly $C_5$, $C_7$ and so on are numerical coefficients).
Numerical lattice evidence suggests \cite{HLE} that $C_3$ is strongly suppressed.
Assuming that only $C_3$ exists,
and fitting the results for the $k$-strings,
Del Debbio, Panagopoulos, and Vicari obtain \cite{HLE}
     $$         C_3= -0.01\pm 0.02\,.$$

At $N=\infty$ and $k$ fixed Eq. (\ref{fexpp})  yields that $\sigma_k = k\, \sigma_1\,.$
Including the next-to-leading term in the $x$-expansion
we get
\beq
\sigma_k = \sigma_{\rm f}\,  k \left(1- \,\, \tilde C_3 \,\frac{k^2}{N^2}\right)\,.
\label{sotix}
\eeq
The term $\sigma_{\rm f}\,\tilde C_3 \, k^3/N^2$ can be interpreted as the binding energy.
The $1/N^2$ factor is well understood, see Ref. \cite{5}.
A challenging question is to understand
why the  next-to-leading correction to 
$\sigma_k$ scales as $k^3$. Alternatively, one can say that
the binding energy per one fundamental string in the $k$-string compound
scales as $k^2$. Experience based on numerous discussions with our colleagues
tells us that the first
 guess is, rather, that it is  the first power of $k$ that must appear
in the binding energy per one fundamental string.  

To get a better idea, let us turn to the supersymmetric $k$-wall tension
where the exact result is known.
In this case the $k^3$ dependence is very general and is a consequence of the
$Z_N$ symmetry (see Appendix). Apparently, the $Z_N$ symmetry 
is also instrumental for $k$-strings.
For the U(1) $k$-strings (i.e. the Abrikosov-Nielsen-Olesen vortices),
the $k$ behavior of $\sigma_k/k$ is expected to be  different.

A physical picture underlying string dynamics that explains
Eq. (\ref{sotix}) is presented in Sect. \ref{ppbsf}.

\subsection{How exact is  the sine formula (in the saturation limit)?}
\label{exactsine}

\noindent

The tension of the $k$ string, $\sigma_k$, a crucial
parameter
of the confinement dynamics,  is under intense scrutiny since the
mid-1980's.
Most frequently were discussed  two competing hypotheses:
(a) the Casimir scaling and (b) the  sine formula originally suggested
by Douglas and Shenker \cite{Douglas:1995nw}
(for extensive reviews and representative list
of references see e.g. \cite{two,three}). The Casimir scaling hypothesis
was recently shown 
\cite{5}
to be inconsistent with the large $N$ expansion.
In the same paper we argued that the 
 sine  formula should hold; in our terms the 
 sine  formula can be presented as follows:
\beq
f(x) = \frac{\sin x}{x}\,.
\label{dsf}
\eeq
It amounts to keeping only the first term in the 
general expansion (\ref{fexpp}). The suppression of higher terms
in this expansion cannot be inferred on general grounds alone;
it is a dynamical statement following from the model \cite{5}.

In the previous section we wrote down the most general expression
for the $k$-string tension \eqref{fexp}, that is compatible with the
$Z_N$ symmetry of the problem and $1/N^2$ nature of
corrections. In the saturation limit
the expression simplifies further to \eqref{fexpp}.
It is tempting to speculate that the first term, the sine, controls
the dynamics in the saturation limit \eqref{satli}, 
namely, that the $k$-string tension
is exactly given by
\beq
 \sigma _k = \Lambda ^2 N \sin \pi {k \over N}. 
\label{esine}
\eeq 

While there is no proof that the relation \eqref{esine} is exact, 
there are arguments suggesting that it might be exact or, at least,
present a good approximation (although we hasten to
add that a parameter controlling this approximation is yet to be discovered).
Below we will summarize theoretical
evidence in favor of the above assertion.

The QCD string is not obviously a BPS object in \None gluodynamics.
Therefore, in principle, it could get higher order corrections in
powers of sine, as in \eqref{fexpp}. The statement that Eq.~\eqref{esine} is exact
is equivalent to the statement that the QCD-string approaches a `` BPS status''
in the saturation limit. Below we explain why a BPS QCD string should
satisfy a sine relation. An additional general argument is given in
the Appendix.

The first hint in favor of the sine formula comes from the large-$N$
analysis of \Ntwo SYM theory softly broken to \None SYM by the adjoint
mass term $m$, due to Douglas
and Shenker \cite{Douglas:1995nw}. In slightly broken \Ntwo SYM the 
QCD string is BPS-saturated
to the leading order in $m$; therefore, no surprise that the sine formula
$ \sigma _k = m\Lambda  N \sin \pi {k / N}$ was found. Details are
as follows. In the softly broken \Ntwo theory Douglas and Shenker found
that
\beq
\sigma_k \sim m N^2 M_{k,k+1},
\eeq
 where $M_{k,j}$ are the masses of the BPS $W$ bosons, 
 $$
 M_{k,j}=| m_k-m_j|\,.
 $$
Here
$$
m_k \sim \Lambda N \left(\sin {\pi k \over N} - \sin {\pi (k-1) \over N}
\right)\,.
$$
 At large $N$, to the leading order in $1/N$, (keeping
 $k/N$ fixed), one then arrives at the sine formula,
 $$
 M_{k,k+1} \sim {\Lambda \over N} \sin {\pi k \over
N}\, ,\qquad \sigma_k \sim m \, \Lambda \, N\,  \sin {\pi k \over
N}\,.
$$ 

At order $m^2$ the strings cease to be BPS, and, accordingly,
the $m^2$ correction to the string tension
was found~\cite{KKon} to defy the sine formula.

The second evidence comes from MQCD. Here again the sine formula is obtained
\cite{Hanany:1997hr}, without an {\em a priori} reason.
Though the MQCD theory is not
QCD, it is not clear why the obtained result is the exact sine. A
possible reason is that MQCD possesses more symmetries than \None SYM,
but no such a symmetry is known at present.

The third hint is the derivation of the $k$-string tension
via supergravity \cite{Herzog:2001fq}. Here the result is model-dependent.
For the Klebanov-Strassler background the sine formula was found
to be an excellent approximation, but not exact. It was found
to be exact for the Maldacena-Nu\~{n}ez background. While both backgrounds
are conjectured to be a dual description of
 pure \None gluodynamics in the far infrared, they have  different 
ultraviolet
contents. For this reason, presumably, different results were obtained.
Note that the above arguments do not apply to  pure \None SYM theory.
Moreover, only the first argument is valid within field theory {\em per se}.

The last argument in favor of the sine formula --- a field-theoretic one
---  is the relation between
the $k$-string tension and the $k$-wall tension in \None gluodynamics
\cite{5}. It is known that the BPS domain wall tensions in \None gluodynamics 
are exactly given by the formula  
\beq
T_k = N^2 \Lambda ^3 \sin \pi {k\over N}.
\eeq 
If one accepts the picture advocated in \cite{5}, that domain walls
are built from a network of
QCD-strings connected by baryon junctions, one 
immediately arrives at the sine formula, for a detailed
 discussion see Ref.~\cite{5}. In this picture, the QCD-string 
effectively becomes a BPS object
in the saturation limit.
The suggestion of the ``walls built of the string network''
is admittedly a model, albeit motivated by various arguments.
The most crucial question we see at the moment
is whether the expansion parameter controlling the accuracy of this model
is numerical or related to some well-defined limits,
such as large $N$, saturation limit, and so on. In the absence of the analytic answer 
to this question one may resort to lattices, see the remark
after Eq.~(\ref{fexpp}).
While the saturation limit is clearly difficult to achieve on the lattice,
the effort is worthwhile.

\section{Decays of quasi-stable strings in the saturation limit}
\label{dqsssl}

\noindent
In Sect.  \ref{ks}
we have discussed the decays of the quasi-stable $k$-strings
with a relatively small $k$, i.e. $k\sim N^0$, for instance,
those attached to the source
$Q^{\{i_1 i_2\}}$.
In this section we will deal with a generic representation
obtained from the
probe color sources of the type
$Q^{i_1,i_2,..., i_k}$ or $Q_{i_1,i_2,..., i_k}$,
with $k$ fundamental or anti-fundamental
indices and $k\sim N$.
 This will correspond to $x$
not close to zero or $\pi$ in the consideration of Sect~\ref{ksslsf}.

The reducible representation $Q^{i_1,i_2,..., i_k}$ can be decomposed
in a number of irreducible representations. For $k<N$
 the number of different types of representations of SU($N$) equals to\,\footnote{We
are grateful to Haris Panagopoulos for pointing out to us this formula.}
$P(k)$ where  $P(k)$  = number of partitions of the integer $k$,
\beq
P(k) \to \exp(\pi \sqrt{2k/3}) / (4 k \sqrt 3)\, , \quad k \gg 1\,.
\label{pk}
\eeq
All strings except that attached to the fully anti-symmetric
sources are quasi-stable. Their tensions are expected to cover densely
the shaded area in Fig.~\ref{defect}. The tensions of ``neighboring''
strings may be expected to differ by an exponentially small amount.
The decay rate of a given quasi-stable string into a
certain
 close-lying string will be enormously suppressed. The density of final states is
exponentially high, however. We are interested in inclusive probability,
i.e. the probability of decay of   a given quasi-stable string
into any state, with the possible subsequent cascading into the stable
anti-symmetric string.

To do the estimate we will ``fuse'' possible final states
into one {\em effective} final string (say, anti-symmetric), 
ignoring the fact that the string tensions
are spread practically continuously.
We will estimate the decay rate by evaluating 
a direct single-leap  tunneling into
the effective final string. 

In Sect. 3.4 we explained that
the conversion of the symmetric two-index string into
the antisymmetric one can be considered
as a quasiclassical tunneling process which, in turn, can be treated
through ``bubble formation" \cite{Voloshin:1995in}. It was important
that the tension difference between the initial and final strings
was suppressed by $1/N^2$ while the mass of the gluelumps
produced was not suppressed by $1/N$ factors.
Under these conditions the approximation of thin domain wall \cite{Kobzarev:1974cp,Voloshin:1995in} is
applicable, the exponential factor in the decay rate
does not depend on dynamical details, and is fully
determined by two parameters: the tension difference and the gluelump
mass.

Now we want to extend our estimates to cover the case of $k$ indices
where $k$ is specified in Eq. \eqref{satli}. As we will see, again
the thin-wall bubble formation is {\em the} adequate
approximation. This method guarantees  a
correct evaluation of the exponent in the exponential factor
determining the {\em inclusive} decay rate of a given quasi-stable string.
The inclusive summation over the final state is a built-in feature of the
method. Particular details of the summation affect only the
pre-exponential factor and are, thus, subleading.

A typical decay of a given $k$-string is characterized by a rearrangement
of many indices; in the saturation approximation the number of indices
rearranged scales as $N$. Then the tension deficit
between the initial (quasi-stable) string and a typical decay product
is of the order of unity. This implies, in turn,
that we must set ${\cal E}\sim k\Lambda^2\sim N\Lambda^2$
(for the definition of ${\cal E}$ see Sect. 3.4).

The above rearrangement is accompanied by the production
of $\sim k$ gluon lumps, so that
$T\sim k\Lambda \sim N\Lambda$. As a result, the decay rate
of a generic quasi-stable $k$-string
turns out to be of the order of
\beq
\Gamma_{\rm gen} \sim \Lambda^2 \, \exp (-\gamma\, N )\,.
\eeq
The suppression is still
exponential but less severe than in Eq. (\ref{expdec}).
This enhancement of the decay rate is in one-to-one
correspondence with the enhancement of the binding energy in
the saturation limit $k/N=$ const, $N\to\infty$.

\section{Oversaturated $k$-strings}
\label{osks}

\noindent
For definiteness we will assume in this section the number of colors $N$ to be even.
(Consideration of odd $N$ can be carried out in parallel.)
Let us define
\beq
k_* = \frac{N}{2}\,.
\label{kstar}
\eeq
The stable $k$-string tension grows as a function of $k$, reaching its maximum
at $k=k_*$. It is clear that for $k>k_*$
it is energetically expedient to split an $N$ 1-string cluster
and  annihilate it, remaining with an ensemble
of $(N-k)$ 1-strings.
This is a tunneling process, however.
Quasi-stable strings with $k=k_* + \ell$ do exist ($\ell >0$).
Our task here is to evaluate their lifetime with respect to the decay
into stable strings with  $k=k_* - \ell$.

For simplicity let us start from the probe source
\beq
Q^{[i_1 ... i_k]}\quad\mbox{with}\quad k=k_* +1.
\label{kstarone}
\eeq
The string created by this source will be composed of $k_*+1$ 1-strings,
and will decay into the stable $k$-string with $k=k_* -1$.
How does this decay occur?

\begin{figure}[H]
  \begin{center}
\mbox{\kern-0.5cm
\epsfig{file=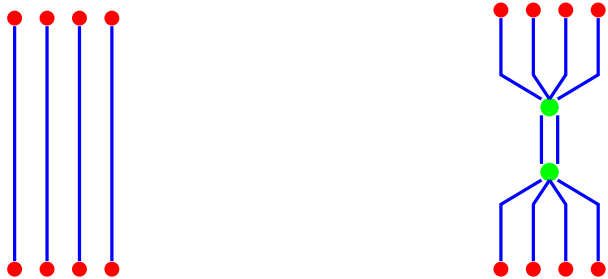,width=8.0true cm,angle=0}}
  \end{center}
\caption{The decay of a 4-string into a 2-string via  creation of
the baryon junctions in SU(6) gauge theory.}
\label{baryon}
\end{figure}

This process is schematically depicted in Fig. \ref{baryon} where we took $N=6$.
Two $N$-string junctions (baryon vertices) are produced
due to tunneling, converting a part of the string between them from
$k= k_*+1$ into $k= k_*-1$. An estimate of the decay rate can be obtained from
exactly the same consideration as in Sect. \ref{ks}. 
The tunneling occurs through a bubble creation. The $N$-string junction scales
as $N\,\Lambda$; therefore, $T\sim N\,\Lambda$, see Eq. (\ref{expdec}).
At the same time the gain in the ``energy density''
(i.e. ${\cal E}$ in Eq. (\ref{expdec})) is equal to $2\,\sigma_1 \sim \Lambda^2$.
As a result,
\beq
\Gamma [(k_*+1)\to (k_*-1)] \sim \exp \left(-\gamma N^2
\right)\,,
\label{kstaronetwo}
\eeq
where $\gamma$ is a positive numerical coefficient of the order of unity.

In fact, there are reasons to believe that the decay rate
will be further strongly suppressed by a large transverse size of the $k$-string
(see Sect.~\ref{ppbsf}), but we will ignore
this effect for the time being.
It
will be discussed elsewhere.

The same estimate (\ref{kstaronetwo}) is valid for $\ell >1$,
but not scaling with $N$.

\section{Physical picture beyond the sine formula for stable
$k$-strings}
\label{ppbsf}

\noindent
In Ref.~\cite{5} we suggested a dynamical model leading to the
sine formula for the stable
$k$-string tensions. Irrespective of this model, the argument of Sect.
\ref{kssl}
shows that the expansion of the $k$ string tension
has the form (\ref{sotix}). While the $1/N^2$ factor in the first  correction
is perfectly clear, it is instructive to get
a physical understanding of the $k^3$ scaling of the
$1/N^2$ correction.

\begin{figure}[H]
  \begin{center}
\mbox{\kern-0.5cm
\epsfig{file=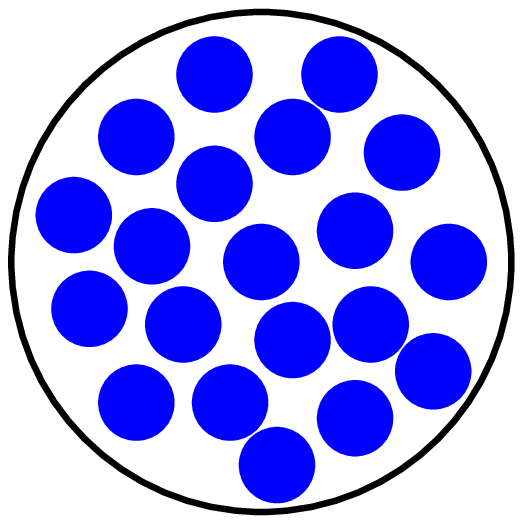,width=3.0true cm,angle=0}}
  \end{center}
\caption{A slice of a $k$-string.}
\label{slice}
\end{figure}

First of all let us note that at $N\gg 1$ the 1-string interactions are weak,
and a quasiclassical picture is applicable. In order to reconcile the facts
that both the flux of the $k$-string and its tension (in the
leading order) grow as $k$ we have to conclude that
the total transverse area of the $k$-string scales as $k$ at $k\gg 1$,
see Fig. \ref{slice}.
Thus, the $k$-string presents an ensemble of loosely bound 1-strings,
with $k$ non-intersecting cores of 1-strings. The interaction occurs only
at the periphery, in the gaps. This is very similar to the structure
of molecules.

There are of the
order of $k$ gaps altogether, each having the area of the order
of $\Lambda^{-2}$. Let us compare the chromoelectric fields at the periphery of the
1-string core in two cases: (i) isolated 1-string; (ii)
1-string which is bound inside the $k$-string.
It is clear that the distortion of the field at the periphery 
in passing from (i) to (ii)
is of the order of $k/N$. The energy per one gap
(which is proportional to the chromoelectric field squared)
thus scales as $(k/N)^2$. Given that the number of the gaps scales as
$k$, we naturally obtain the $k^3/N^2$ regime
for the first correction in the $k$-string tension. 

Within this picture the expansion parameter $(k/N)^2$,
with the overall $k$ factor, occurs naturally, leading to the
scaling parameter $x$ introduced in Sect. \ref{kssl}.

The linear growth of the string cross section with $k$ at large $k$
is a prediction following from our analysis.

We check the above picture in a toy supersymmetric
model in the Appendix.

\section{Tension defect vs. the scaling variable \boldmath{$x$}}
\label{tdvsvs}

\begin{figure}[H]
  \begin{center}
\mbox{\kern-0.5cm
\epsfig{file=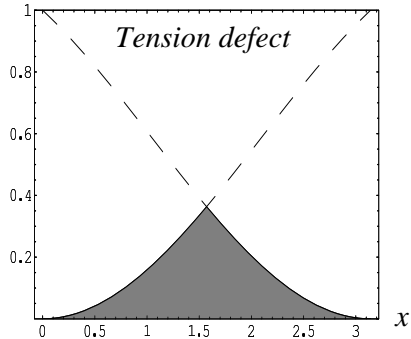,width=7.0true cm,angle=0}}
  \end{center}
\caption{The tension defect. The two dashed curves in the upper
side of the plot present oversaturated strings.}
\label{defect}
\end{figure}

\noindent
In this section we will discuss the binding energy
of stable and quasi-stable $k$-strings in the saturation limit.
It is convenient to introduce a quantity
analogous to the mass defect in nuclear physics.
We will call it {\em tension defect} $\,\, {\cal T}_k$,
\beq
{\cal T}_k = 1 - \frac{\sigma_k}{\sigma_{\rm f}\, {\rm Min}\{k,\,\, N-k\}}\,.
\label{tendef}
\eeq
If the sine formula is correct for the stable $k$-strings
(up to $1/N$ corrections), then
\beq
{\cal T}_k = 1- \sin x \,\times\, \left\{ \begin{array}{l}
x^{-1}\,\,\, \mbox{at}\,\,\, x < \pi /2\,,
\\[2mm]
(\pi -x)^{-1}\,\,\, \mbox{at}\,\,\, x > \pi /2\,,
\end{array}
\right.
\label{curve}
\eeq
see Fig. \ref{defect}. The presence of higher powers of sine in Eq. (\ref{fexpp})
may somewhat change the precise form of the curve, but qualitatively its
shape should remain as in Fig. \ref{defect}.
For stable strings, the maximal tension defect is
$1- (2/\pi )$, achieved at $x=\pi /2$.

The area between the curve (\ref{curve})
and the horizontal axis is covered by quasi-stable strings. 
Indeed, for the given $N$-ality $k$ we have
$P(k)$ distinct representation, see Eq. (\ref{pk}), starting from the fully
anti-symmetric, through mixed symmetries, up to
fully symmetric. $P(k)$ is exponentially
large. We expect that the tension splittings are roughly
the same. Then it is obvious that the string tensions
of the quasi-stable strings are separated from each other
by exponentially small intervals. 

Under the circumstances it would be appropriate
to replace the discrete tension spacings by a continuous distribution
$\rho ({\cal T}; x)$, which for each given $x$
gives the density of strings  per interval $d {\cal T}$.
Calculating $\rho ({\cal T}; x)$ is a very interesting dynamical problem.

\section{More on the Casimir scaling}
\label{mocs}

In Sect.~\ref{mrllks} we  mentioned that: (a)
many lattice measurements detect quasi-stable strings, with no hint on
their breaking; and (b) observe the Casimir scaling in the ratio 
of the tension of the given string to 
that of the fundamental string. For instance, in Ref.~\cite{Deldar}
ample data are presented regarding the adjoint and a number of
other quasi-stable strings in SU(3). No deviations from the Casimir 
scaling are seen withing
the achieved accuracy. 

According to the Casimir formula,
\beq
\sigma_{\rm adj} / \sigma_{\rm f} = 9/4\,,\qquad (\mbox{SU(3) Casimir})\,.
\eeq
The adjoint string is built of two fundamental ones,
and at distances $\gsim \Lambda^{-1}$ they attract, so
that placing these two strings at an appropriate distance
one must and can get the above ratio less than 2. Even if we take the 
separation distance between the two 1-strings involved
very large, the ratio will equal 2.

How can one get this ratio larger than 2?

Typical string sizes on the lattice are $\lsim 2$ fm, while the transverse size
of the fundamental string is $\sim 0.7 $ fm.
As we have argued above, the transverse size of the
composite string must be even larger. Thus, the length/thickness ratio is in fact
not large. 

\begin{figure}[H]
  \begin{center}
\mbox{\kern-0.5cm
\epsfig{file=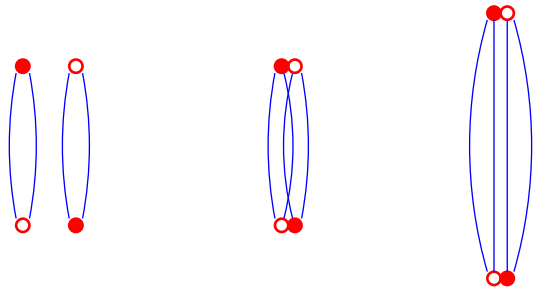,width=7.0true cm,angle=0}}
  \end{center}
\caption{Various flux tubes. a. Two separated fundamental strings. b.
A short adjoint string. c. A long adjoint string.}
\label{mocsf}
\end{figure}

If one could measure the tension of the configuration
depicted in Fig.~\ref{mocsf}a, because of the attraction, one would
get less than $2\sigma_{\rm f}$. The only way to get more than 
$2\sigma_{\rm f}$ is to force the fundamental strings to overlap.
Then, because of the repulsion of the overlapping fluxes,  (prior to the string decay),
see the end of Sect.~\ref{prel},
one will get 
$> 2\sigma_{\rm f}$ for the energy per unit length.

This is precisely what happens if one considers the probe quark of the form
$Q_i^j$, see Fig.~\ref{mocsf}b. Because of the insufficient
separation between $Q_i^j$ and $\overline{Q_k^l}$ the overlap has to be
substantial. For {\em bona fide} long strings
(Fig.~\ref{mocsf}c) the overlap would be insignificant,
and will affect only the part of the potential which does not scale with $L$
as $L^1$.

Still, the question remains: Why a sausage-like configuration (Fig.~\ref{mocsf}b),
which develops at intermediate distances,  is characterized by energy per unit length
which follows the Casimir formula?

Here we suggest a tentative ``semi-empiric'' answer.
It is known since the 1970's \cite{svz}
that basic properties of low-lying quark mesons\,\footnote{
By low-lying we mean the lowest radial excitations
in each channel with the given spin and parity.} 
are to a large extent determined by the gluon condensate.
The coefficient in front of the gluon condensate is proportional
to the quadratic Casimir operator. Therefore, as long
as the higher condensates are not very important numerically,
the masses of the lowest mesons will be related to the 
 quadratic Casimir operators. These latter masses, in turn, 
are determined by ``short strings.''

If this explanation is correct, approximate Casimir scaling
should take place at intermediate distances
when the overlap between constituents of the composite string
is significant.

\section{Orientifold theories vs. supersymmetric gluodynamics}
\label{otvsg}

\noindent

The orientifold gauge theory is obtained from
SU($N$) supersymmetric gluodynamics by twisting
the color indices of the fermion field.
The fermion sector of the orientifold gauge theory
consists of two Weyl spinors, in the color representations
$$\asymm\quad \mbox{and}\quad  \basymm\,\, .$$
One can combine them in one Dirac spinor. Recently it was shown 
\cite{asv} that
in the planar limit $N =\infty $ the orientifold gauge theory
is equivalent to ${\cal N}=1$ supersymmetric gluodynamics in its
bosonic sector (For a general introduction to the
topic of planar equivalence see Ref.~\cite{matt}.)

In connection with the $k$-string tensions,
the above equivalence {\em apriori} might seem rather puzzling.
Indeed, the only dynamical fields in supersymmetric gluodynamics
are those belonging to the adjoint representation ---
they can screen no probe color charge with a non-vanishing $N$-ality.
At the same time, the anti-symmetric two-index ``quarks'' of the
orientifold theory do screen probe color charges with
the even $N$-ality, 2, 4, 6, etc. This seemingly results in 
distinct dynamical patterns.

In fact, there is no paradox. Let us remember that the above equivalence
takes place only in the limit  $N =\infty $. 
Non-planar effects are certainly different in the supersymmetric and
orientifold (parent and daughter) theories.
Breaking the string --- that's how the probe color-charge screening
is seen in the hadronic language --- is a non-planar effect.

Thus, we expect then the Wilson loops of the parent and daughter theories
to be related, only provided  their area, being large compared to $\Lambda^{-2}$,
does {\em not}  scale with $N$. In this case, as we already know from the previous
sections (see also Sect. \ref{impli}), there is no time for screening to occur
in either theories. We hasten to add, however,
that for such contours the accuracy of the measurement
of the $k$-string tension is such that sub-leading in
$1/N$ corrections cannot be detected,
and we will arrive at a rather trivial formula
\beq
\sigma_k = \kappa\,\sigma_1\,,
\label{trivfor}
\eeq
where 
$\sigma_k$ is the $k$-string tension, $\sigma_1 \equiv \sigma_{\rm f}$
and the parameter $\kappa$ is introduced in Eq. (\ref{kap}).
Detecting $1/N$ corrections on the right-hand
side of Eq. (\ref{trivfor}) would require contours with area growing with
$N$, which would invalidate the planar equivalence.

\section{Implications of our results for lattice measurements}
\label{impli}

\noindent
Throughout the paper we discussed the decays of quasi-stable strings
into stable strings within the Minkowskian picture. 
Lattice measurements are mainly performed within the
 Euclidean approach. Most commonly
lattice theorists measure Euclidean Wilson loops or
correlators of the Polyakov lines (see e.g. Ref.~\cite{Greensite:2003bk}).
It is natural to ask what is the impact of our results on the Euclidean 
lattice measurements. Below we will discuss two typical examples.
Other examples can be treated in a similar manner.

\subsection{Decay of the adjoint string from the Euclidean standpoint}
\label{dases}

Let us  first discuss the adjoint string decay  into a broken
string. In Sect.~\ref{as} we  estimated the life-time of the adjoint string
as $\tau \sim \Lambda ^{-1} N^2$. When one studies the adjoint string
by calculating the Euclidean Wilson loop (in the adjoint
representation), one arrive at a two-term formula\,\footnote{We are grateful
to J.
Greensite for an insightful discussion of this issue. See also 
Ref.~\cite{Greensite:2003bk}.}
\beq
\langle {\cal W}\rangle\,  \sim\,  e ^{-\sigma_{\rm adj} {\cal A}} + {1\over N^2}
\,  C \, e ^{-\mu P}\, ,
\label{2term}
\eeq
where the numerical coefficient 
$C$ is $N$ independent but is very small (see the end of Sect.~\ref{fsiqcdwdq}
for the relevant discussion). The suppression of $C$ 
is determined by geometrical factors, the appropriate parameter being
$L\Lambda$. The numerical parameter $\mu $ in the second term in Eq.~(\ref{2term})
is of the order of $\Lambda$. In fact, it represents the gluelump mass,
which is expected to be rather large, in the ballpark of 0.8 to 1 GeV.
It is quite obvious that for the asymptotically large
contours the perimeter term (representing the adjoint string breaking)
always wins. However, given the pre-exponential suppression and a
relative largeness of $\mu$, at intermediate sizes there
 will be a competition between the area term and
the perimeter term.

The perimeter behavior is non-planar and, therefore, it is suppressed by 
$1/N^2$. In addition, as was mentioned,
there is an additional suppression due to the fact that
the final state producing the perimeter
law is {\em not} a state with  highly excited mesons but, rather,
that with the lowest-mass
 mesons. By
equating the area term with the perimeter term we arrive
at an estimate for the (Euclidean time) duration of the process \cite{DelDebbio:1992yp} 
\beq
T \sim \Lambda^{-1}\ln \left(  N^2 /C \right) \,, \qquad C\ll 1\, .
\label{adjet}
\eeq 
The critical contour size
needed to see the adjoint string breaking grows with $N$, albeit rather slowly.

\subsection{Decay 
(conversion) 
of the excited \boldmath{$k$}-string from the Euclidean standpoint}
\label{deses}

The prime emphasis  in this paper was on the decay of the excited
$k$-strings. Let us consider for definiteness 2-strings ---
symmetric and anti-symmetric. The Minkowskian description of the
decay of the symmetric string into the anti-symmetric one
is presented in Sect.~\ref{ks}. Here we will discuss the impact of our result
on the Euclidean measurement of the Wilson loop in the symmetric 
two-index representation
of SU($N$).

The two-term formula (\ref{adjet}) now has to be modified,
since, after the breaking of the symmetric string,
the anti-symmetric one is formed and, therefore, both terms
 exhibit the area law,
\begin{eqnarray}
\langle {\cal W}\rangle\,&\sim &     e^{-\gamma N^2}\,C\, 
e^{-\sigma_{\rm anti-symm} \, {\cal A}} +   e^{-\sigma_{\rm symm}\, {\cal A}}
\nonumber\\[2mm]
&=& e^{-\sigma _{\rm anti-symm} \, {\cal A}}\left\{ C\, 
e^{-\gamma N^2} + e^{-\Lambda^2\,{\cal A}/N^2}\right\}\,.
\label{2termk}
\end{eqnarray}
 As was explained in Sect.~\ref{ks}, the transition into the stable (anti-symmetric)
state is
exponentially suppressed, since it is a tunneling process.
It is also suppressed by a numerical (geometrical) factor C ($C\ll 1$) which
is $N$-independent. In the second line of Eq.~(\ref{2termk}) we use the fact that
the tension difference is 
\beq
\sigma _{\rm symm}-\sigma _{\rm anti-symm} = \Lambda^2/N^2\,.
\label{delsas}
\eeq

 To guarantee that the 
first term wins (i.e. the decay of the symmetric symmetric 
string into anti-symmetric becomes visible)
one needs the area
\beq
{\cal A}_{\rm crit} \gsim \Lambda^{-2}\left\{\gamma N^4 -(\ln C ) N^2
\right\}\,,\qquad C\ll 1\,.
\label{kareac}
\eeq
The numerical value of the coefficient $\gamma$
is unknown. Given the extremely steep $N^4$ dependence of ${\cal A}_{\rm crit}$
it will be extremely difficult to measure the symmetric  string conversion
into the anti-symmetric  one  within the existing
 lattice methods, unless $\gamma$ is abnormally small, which, {\em apriori}
seems quite unlikely. 

Equation (\ref{delsas}) implies that the area necessary for measuring
$\sigma_{\rm symm}$ (more exactly, the split between $\sigma_{\rm symm}$
and $2 \sigma_{\rm f}$)
must scale as $\Lambda^{-2} N^2$. This is especially obvious
if the measurement is carried out through
 the Wilson loop in the reducible two-index representation,
or $\langle {\cal W}_{\rm fund}^2\rangle$.

\section{Discussion and conclusions}
\label{dac}

\noindent
In this paper we started developing a systematic
description  of quasi-stable and stable composite 
QCD strings ($k$-strings) based on the $1/N$ expansion.
The paper consists of three main parts. First, we considered
various quasi-stable  strings and found their decay rates
(per unit time per unit length). These results are summarized in Table \ref{table1}.
The picture that we developed here is  the physical Minkowskian
one. We have viewed the quasi-stable strings in their
relation to  quasi-stable mesons
that can break due to non-planar interactions. The key observation was
that at large $N$, the probability is small and hence the 
life-time of the string (and, correspondingly, excited  resonances) is large.

\begin{table}[H]
\begin{tabular}{|c|c|c|c|c|c|}
\hline
& {\small\em Fundamental} & {\small\em Adjoint} & {\small\em Excited} & {\small\em Excited} 
& {\small\em Over-}
  \\
& {\small\em string} & {\small\em string} & {\small\em $k$-strings} & {\small\em $k$-strings} &
 {\small\em saturated}
  \\
& & & & & {\small\em strings} \\
& {\small QCD with} & {\small Pure YM} & {\small $k$ small} & {\small Saturation} &
  \\
& {\small dynamical quark} & {\small or SYM} &  & {\small limit} &  \\
\hline\hline
& & & & & \\
$\Gamma = \frac{W}{ \tau L}$ & $\frac{\Lambda ^2}{ N}$ & $\frac{\Lambda ^2}{ N^2}$ &
  $\Lambda ^2e^ {-\gamma N^2}$ & $\Lambda ^2e^ {- \gamma N}$ & $\Lambda ^2e^{ -\gamma N^2}$ \\
& & & & & \\
\hline
\end{tabular}
\caption{Decay rates (per unit length per unit time) of
quasi-stable strings.}
\label{table1}
\end{table}

Intuitively it is clear that the longevity of the quasi-stable strings
at large $N$ must imply that detecting the breaking of
such strings in the Euclidean lattice simulations
must require large areas. In fact, this is the {\em raison d'etre}
of the adjoint strings, of the symmetric strings with
the tension distinct from that of anti-symmetric, and so on.
Numerous observations of these quasi-stable strings at ``intermediate
areas''
are reported in the lattice literature, with very few (if at all)
observations of the string breaking.

The central question is --- what does it mean, an ``intermediate area''?
To answer this question we need to connect
 our Minkowskian picture to the 
Euclidean approach based on lattice measurements
using rectangular  Wilson loops or Polyakov line correlators. 
In Sect.~\ref{impli} --- our second main part ---
we gave estimates of the contour areas needed
in order to detect the string breaking (conversion).
The $N$ dependence of the critical area for the adjoint string is logarithmic.
In contrast, to see the conversion of excited $k$-strings into anti-symmetric
one would need to deal contour areas growing as powers of $N$.

Finally, the third  main part of this paper is devoted to the tensions of
the composite strings. 
We summarized old and presented some new arguments
in favor of the sine formula for the tensions
of the stable (fully anti-symmetric)
$k$-strings. We made estimates regarding the splittings
between the tensions of the quasi-stable strings and the tension of the
stable string. We introduced the notions of the tension defect,
saturation limit, and oversaturated strings. We tried to explain
why the Casimir scaling which in many instances predicts
$\sigma_{R}/\sigma_{\rm f} >\kappa $, where $\kappa $ is 
introduced in Eq. (\ref{kap}),
 is physically senseless, on the one hand, and still,
nevertheless, many existing lattice measurements seem to confirm
the ratio of the Casimir operators for $\sigma_{R}/\sigma_{\rm f}$.
  
Now,  we would like to say a few words  on the actual QCD ($N=3$) string.    
In the actual world $1/N=1/3$, and one may wonder
which aspects of (or conclusions from)
our analysis are applicable
for such not-so-small value of the expansion parameter, if at all.
Besides, for the SU(3) color group,
only one string --- the fundamental one --- is truly stable.

We have seen that in some cases the relevant  parameter is $N^2\sim 10$.  
 Although we do not know numerical coefficients,
it is reasonable to expect that the conclusions based
on this parameter will survive, at least, at a semi-quantitative level.

\Acknowledgements

We thank J. Ambjorn, P. de Forcrand, S. Deldar, 
L.~Del Debbio, J.~Greensite, K. Konishi,
A. Kovner, C. Korthals Altes, B. Lucini, 
M. L\"{u}scher, H. Meyer, H.~Panagopoulos, P. Rossi, M. Teper, and E. Vicari
  for numerous enlightening and fruitful  discussions. Stimulating communications
with D. Khmelnitskii and V.~Shev\-chenko are gratefully acknowledged.
A.A. thanks Imperial College, Oxford University and Niels Bohr
Institute where a part of this  work was carried out,
for warm hospitality.

The work of M.S. is supported in part  by DOE grant DE-FG02-94ER408.

\section*{Appendix}

\renewcommand{\theequation}{A.\arabic{equation}}
\setcounter{equation}{0}

This Appendix can be viewed as a supplement to Sect.~\ref{ppbsf}
where we argued that the $k^3/N^2$ behavior of the $k$-string binding energy
is natural. Here we illustrate this argument by an exactly solvable
model, which exhibits some general features of the
$k$-string ensemble.

Consider s supersymmetric Wess-Zumino model
with one complex scalar field and one complex two-component spinor.
In four dimensions this is the theory of one chiral superfield
with the minimal supersymmetry (${\cal N}=1$, four supercharges),
while in two and three dimensions the supersymmetry is extended.
While the concrete form of the superpotential is unimportant,
it is important to choose it to be $Z_N$ invariant.
An appropriate choice is
\beq
W = \frac{N}{2}\left( \Phi - \frac{1}{N+1}\, \Phi^{N+1}\right)\,.
\eeq
Irrespective of the number of dimensions
(two, three or four), the central charge is
\beq
Z_{n,n+k} = \frac{N}{2}\left[ \exp\left( i \frac{2\pi}{N} (n+k)\right)
- \exp\left( i \frac{2\pi}{N} k\right)
\right]\,.
\eeq 
The mass of the BPS object built of $k$ constituents
is
\beq
M_k = N \sin \frac{\pi k}{N}\,.
\eeq
This is an exact formula.

It is easy to see that the only ingredients important for the derivation are:
(i) the holomorphic nature of the central charge; and (ii) $Z_N$ symmetry.
At $N\to \infty$ and $k\ll N$ the mass defect is
\beq
M_k - k M_1 = -\frac{\pi^2}{6}\frac{k^3}{N^2}\,,
\label{chety}
\eeq
i.e. the same as discussed in 
 Sect.~\ref{ppbsf}. In the case at hand, however, the problem is exactly solvable, 
everything is known, and the question of the physical interpretation is easy.

The BPS object built from $k$ constituents
(i.e. $k$ elementary solitons) has mass (\ref{chety}). Each elementary
soliton is separated from its neighbor by a gap, in which the field
$\Phi$ deviates from its vacuum value only slightly.
It is not difficult  to see that the mass defect is saturated in the inter-soliton gaps, 
due to a modification of the soliton tails far from their cores.
There are $k$ such gaps, and $\delta \Phi $ in the gaps is typically
$\sim k/N$. Then the mass defect produced
by each gap scales as $k^2/N^2$.
 The overall mass defect scales as $k\times k^2/N^2$.

\end{document}